\documentclass[12pt,a4paper]{cibb}

\usepackage{subfigure,graphicx}
\usepackage{amsmath,amsfonts,latexsym,amssymb,euscript,xr}
\usepackage{booktabs}
\usepackage[nodayofweek]{datetime}
\usepackage{hyperref}

\usepackage[table]{xcolor}
\usepackage{color,colortbl,tabularx}

\usepackage[english]{babel}
\usepackage[protrusion=true,expansion=true]{microtype}
\usepackage{amsmath,amsfonts,amsthm}
\usepackage{pifont}

\definecolor{LightBlue}{rgb}{0.88,0.9,0.9}

\title{\Large $\ $\\ \bf Snakemaker: Seamlessly transforming ad-hoc analyses into sustainable Snakemake workflows with generative AI}

\author{\large Marco Masera$^1$, Alessandro Leone$^{1}$, Johannes Köster$^{2}$ and Ivan Molineris$^{*,1}$}
\address{\footnotesize $\ $\\$^1$ Dipartimento di Scienze della Vita e Biologia dei Sistemi and MBC, Università di Torino, Via Accademia Albertina 13, 10126 Torino, Italy.
\footnotesize $\ $\\$^2$ Algorithms for reproducible bioinformatics, Genome Informatics, Institute of Human Genetics, University Hospital Essen, University of Duisburg-Essen, Essen, Germany.\\
\bigskip
$^*$corresponding author
}

\abstract{\small Sustainable data analysis, Generative AI, Snakemake, Reproducibility. \normalsize
\\[17pt]
{\bf Abstract.} Reproducibility and sustainability present significant challenges in bioinformatics software development, where rapidly evolving tools and complex workflows often result in short-lived or difficult-to-adapt pipelines. This paper introduces Snakemaker, a tool that leverages generative AI to facilitate researchers build sustainable data analysis pipelines by converting unstructured code into well-defined Snakemake workflows. Snakemaker non-invasively tracks the work performed in the terminal by the researcher, analyzes execution patterns, and generates Snakemake workflows that can be integrated into existing pipelines. Snakemaker also supports the transformation of monolithic Ipython Notebooks into modular Snakemake pipelines, resolving the global state of the notebook into discrete, file-based interactions between rules. An integrated chat assistant provides users with fine-grained control through natural language instructions. Snakemaker generates high-quality Snakemake workflows by adhering to the best practices, including Conda environment tracking, generic rule generation and loop unrolling. By lowering the barrier between prototype and production-quality code, Snakemaker addresses a critical gap in computational reproducibility for bioinformatics research.}

\begin{document}
\thispagestyle{myheadings}
\pagestyle{myheadings}

\section{\bf Introduction}
\label{sec:SCIENTIFIC-BACKGROUND}

Reproducibility is a cornerstone of scientific research, ensuring that findings can be independently validated and built upon. In bioinformatics, where computational tools and workflows are integral to data analysis, reproducibility is particularly critical. It underpins the reliability of scientific claims, facilitates peer review, and strengthens the foundation for future discoveries. Yet, reproducibility remains a source of issues, due to dependencies on short-lived or rapidly changing tools and to the complexity and low quality of the published pipelines \cite{A}. This lack of reproducibility is a burden to the entire field, as it limits the possibility of reusing previous pipelines in new scenarios \cite{B}. 

F. Mölder et al. emphasized that, beyond reproducibility, transparency and adaptability are essential for the lasting impact of published software \cite{C}. These qualities, that define sustainable data analysis, entail properties of the code that go beyond the simple ability to execute and relate to the overall quality and thoughtfulness of the published code. Similarly, L.P. Coelho describes three levels of scientific software \cite{D}: level-0 for simple, one-off analysis scripts; level-1 for open-source code published with research papers; and level-2 for reusable, adaptable tools. This ties into K. Ferenc et al.’s argument that bioinformatics often neglects sound software practices, driven by a results-first culture where code is a means to publish, not a lasting asset \cite{E}. As a result, researchers typically produce level-0 scripts for experiments and see refining them into level-1 or level-2 tools as a laborious and unnecessary effort. Over time, as these ad hoc tools are reused within groups, technical debt accumulates, making future improvements harder. 

What if some 'enzyme' could break down the 'activation energy' needed to jump from level-0 to level-1 or level-2 software? This paper proposes Snakemaker, a tool that leverages generative AI to offer semi-automatic conversion of unstructured or use-case-specific code into well-defined Snakemake workflows. 

Snakemaker runs non-invasively, letting researchers prototype as usual while supporting the transition to sustainable pipelines. Generative AI analyzes code and converts it into Snakemake workflows, while detecting patterns, generalizing rules, tracking and exporting Conda environments and generating documentation. A graphical and chat interface gives users control and visibility over the conversion process. 

Snakemaker supports two working modes: Shell and IPython notebooks. 

Shell languages like Bash are core tools in bioinformatics, as they offer access to a wide range of bioinformatics tools and excel at chaining them to process large, text-based datasets (e.g., FASTA, FASTQ, VCF). Typically, shell-based pipelines are built by running commands in the terminal and eventually copying them into .sh files for reproducibility. Snakemaker records the user’s terminal activity and converts it into Snakemake workflows. 

The IPython Notebook support enables semi-automatic conversion of notebooks into structured pipelines. While notebooks are popular for prototyping due to their flexibility and convenient integration of code and visualizations, they often grow into unwieldy, monolithic codebases with hard-to-track execution states. This can lead to uncertainty about results and frequent, time-consuming full re-executions. As classic level-0 software, notebooks are ideal for rapid iteration but ill-suited for reuse, collaboration, or long-term maintenance. Converting them into Snakemake workflows improves modularity, clarifies dependencies, and ensures automatic tracking of data and code changes — boosting reproducibility and sustainability.

\section{\bf Data and Methods}
\label{sec:DATA-AND-METHODS}

Snakemaker is developed as a VSCode extension, integrated into the graphical interface of the workspace. VSCode APIs enable fine-grained control over the integrated terminal and the working directory of the user; moreover, the language API allows to prompt the GitHub Copilot models; while Snakemaker provides support for any OpenAI-compatible LLM API, the integrated LLM support is the simplest, ready to use solution for most cases.

This section describes the implementation of the Bash (\ref{sec:DATA-SHELL-TRACKING}) and Notebook (\ref{sec:DATA-NOTEBOOK-EXPORT}) features, and the challenges of using LLMs for workflow generation, along with the solutions devised.

\subsection{ \bf Shell tracking and conversion}
\label{sec:DATA-SHELL-TRACKING}
This section presents the implementation of the Shell tracking and conversion functionality.
VSCode APIs enable tracking of the terminal activity: user commands, their return codes, and the active Conda environments. The recorded activity is processed by the LLM to suggest rule names, identify potential input and output files, and determine whether each command is relevant to the workflow or a one-off action. This relevance flag is essential for filtering out unrelated operations, such as directory navigation or Git commands. Commands can also be inserted manually into the history, although without the metadata. 
The conversion pipeline consists of several steps:

1) \textbf{Contextual integration}. 
Existing Snakemake workflows in the current working environment are provided as context for the model. The LLM tries to mimic the existing formalisms to guarantee some stylistic consistency in the codebase, and to filter out redundant rules. 

 2) \textbf{Rule drafting}. 
Commands in the history and their metadata are fed to the LLM to produce the first version of the Snakemake rules. The model looks for patterns in commands and existing rules to verify the possibility of merging different commands in a single, generic rule, to unroll bash loops into rules using the expand directive, to avoid repeating rules. 

3) \textbf{Configuration}.
A second LLM analysis looks for hardcoded parameters or paths that can be moved to a config file. Existing config files are used as an example and as a pool of parameters that can be used in the rules. Experimental results have shown that separating this step from the previous one results in a significant improvement both in the quality of the generated rules and the config. 

4) \textbf{Post-processing}.
A post-processing step employs regex and finite-state automata to fix some common formatting errors, and then to merge the new rules with the existing workflow. 

5) \textbf{Validation}.
The last step performs iterative validation and correction. The workflow is passed through the Snakemake binary to find issues, which are then fed back to the LLM to find a fix. Experimentally, one or two iterations are sufficient to fix all errors in most cases. For complex, edge-case errors, a step-back-prompting  \cite{I} technique is employed. Here the model is employed to write an analysis of the problem and a fix plan. This plan is then fed back to the LLM and applied. 

\begin{figure}[h]
\vspace{3mm}
 \begin{center}
 \includegraphics[width=0.8\textwidth]{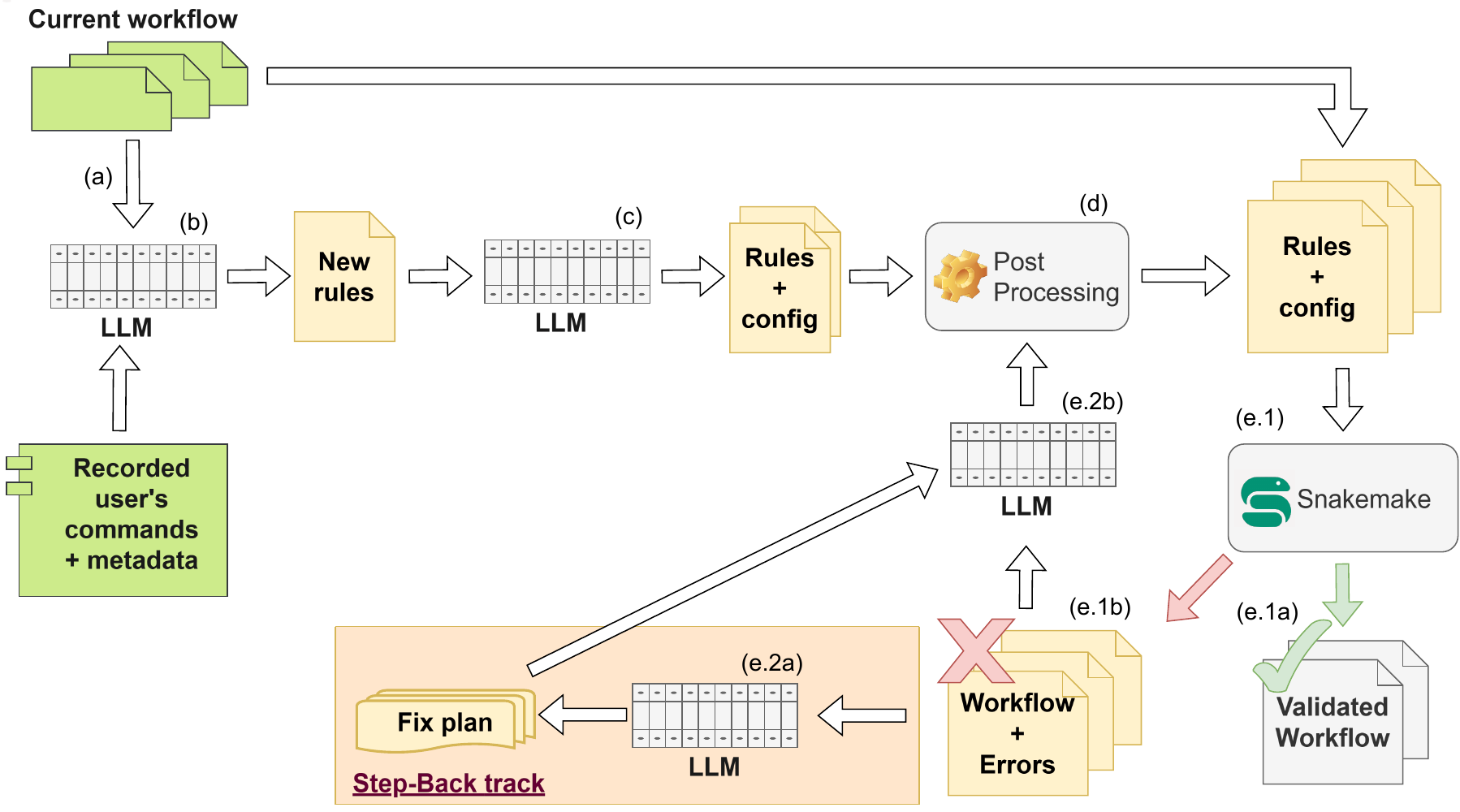}
\caption{\textbf{Bash commands conversion pipeline}.
The current workflow is included in the model's context (a) before generating the new rules (b). A second LLM pass extracts the config (c) and a post-processing step performs some fixes and merges rules with the previous workflow (d). The new workflow is fed to Snakemake (e.1); if errors arise (e.1b) they are either fed to the LLM for direct fixing (e.2b) or follow a two-step pipeline (e.2a). The resulting rule return in the pipeline from postprocessing.
\label{fig:bash}}
 \end{center}
\vspace{-8mm}
\end{figure}

Snakemaker can also generate documentation for the current workflow and the recorded bash history. While the model has no means of knowing the exact intentions of the user, it can take advantage of the comments, the naming of rules and files, and its inner knowledge of the tools to infer the broader scope and produce a description of the implemented workflow. 

\subsection{\bf Notebook conversion}
\label{sec:DATA-NOTEBOOK-EXPORT}
This section presents the implementation of the Ipython Notebook conversion functionality. This feature aims to semi-automatically decompose notebooks into modular Snakemake workflows, and requires a complex unrolling of the global Ipython state into a set of discrete, file-based dependencies between rules. These file-based dependencies are then implemented by enriching cells with additional generated code. 

The global state is unrolled into a directed acyclic graph (DAG), where nodes correspond to cells and edges to data dependencies. The process requires determining, for each cell, the sets of variables that are written and the ones that are read before being written. From these two sets, and an assumption that cells are run in order, the DAG can be built by drawing an edge, for each variable in the read set, from the reader cell to the closest writer of the same variable between cells above the reader. Computing the read and write sets from static analysis is an undecidable problem \cite{J}, but it can be relaxed by always considering the most pessimistic scenarios: if a variable can be read before writing, it is a dependency. 

Function declarations require special handling, as their definitions and calls may appear in different cells. To avoid ambiguity, functions are isolated in dedicated cells and made independent of the global context by renaming and adding to the arguments their dependencies, ensuring the caller is always responsible for providing the data. Import statements are similarly parsed, stripped during pre-processing, and reinserted at export on a per-script basis. 

Identifying the read and write sets has proven to be an error-prone endeavor even for high performance models, with occasional missed writes or hallucinated reads, causing missing dependencies. Following the same \textit{coats of paints} approach of \ref{sec:DATA-SHELL-TRACKING}, a second LLM pass looks for errors in the sets explicitly related to the missing dependencies. 

Each cell is labeled as a rule, script, or undecided. Rules correspond to Snakemake-executed tasks, scripts are reusable code snippets importable by other scripts or rules, and undecided cells need to be labeled manually. Certain constraints apply: scripts can be imported anywhere, but a rule’s output files may only be read by other rules. 

After finalizing the DAG, the dependencies are implemented with generated code. A prefix block implements the input dependencies, reading and parsing files, and a suffix block writes output files for downstream dependencies. Snakemake rules define the connections between cells. As in 2.1, a second LLM pass builds the configuration for the rules.  As in the bash pipeline \ref{sec:DATA-SHELL-TRACKING}, before export iterative validation and correction are performed for both the rules and the Python scripts.  

\begin{figure}[h]
\vspace{3mm}
 \begin{center}
 \includegraphics[width=0.8\textwidth]{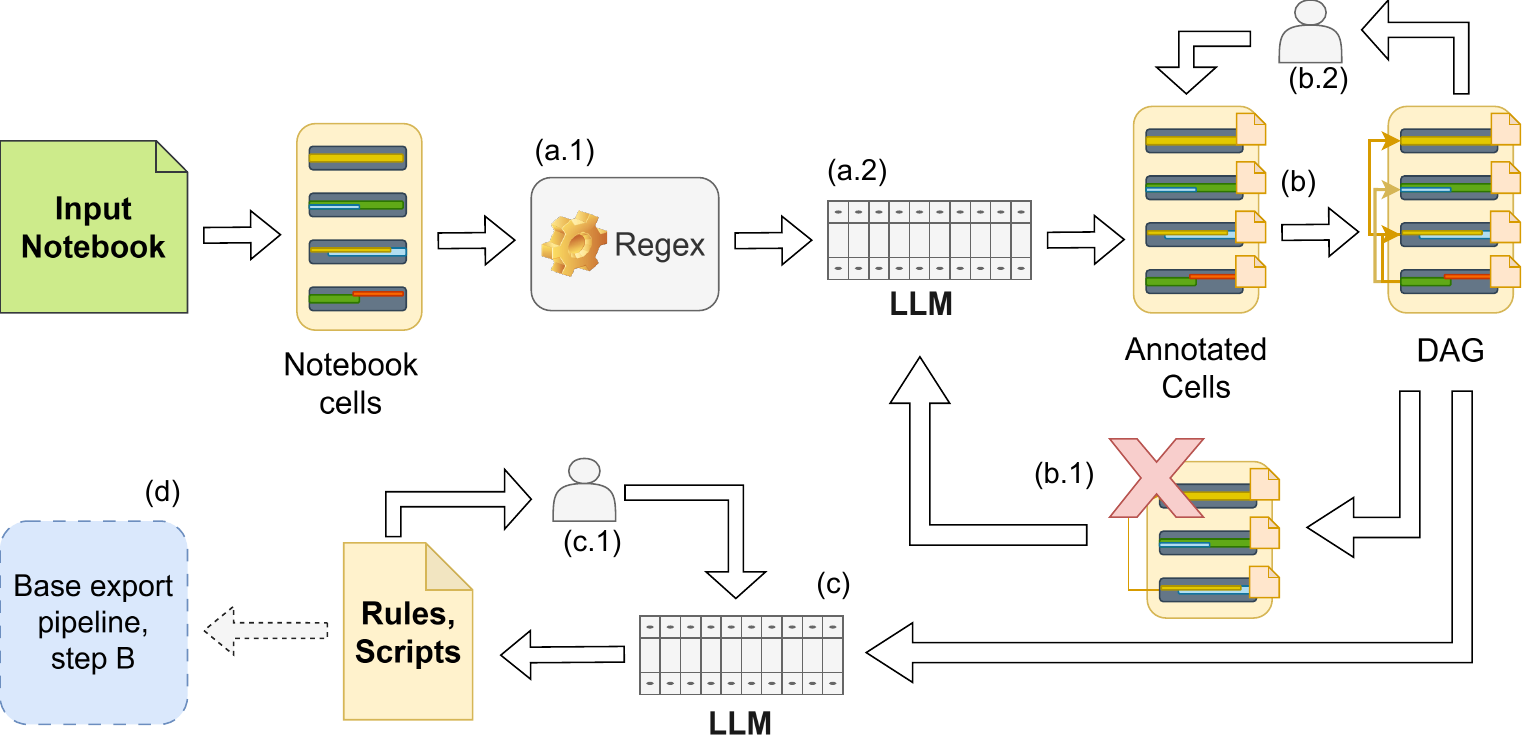}
\caption{\textbf{Notebook conversion pipeline}.
The input notebook is parsed with regex and the LLM (a.1, a.2) to transform and annotate cells with \textit{read} and \textit{write} sets and cells' states. The DAG is built from the annotated cells (b), and in case of errors a second LLM pass is performed (b.1). The user can modify the cells' annotations (b.2) and consequently the DAG. The DAG is then fed to the LLM to generate additional code blocks and Snakemake rules (c); the user can manually change them (c.1), in which case the LLM pass is repeated to propagate changes. The set of rules and scripts can then be fed to the conversion pipeline described in figure \ref{fig:bash} from step b.
\label{fig:notebook}}
 \end{center}
\vspace{-8mm}
\end{figure}

\subsection{\bf Chat Assistant}

The functionalities described above are based on some designed, structured way in which the extension is supposed to operate with minimal user intervention. The chat assistant feature, on the other hand, is meant to be the Swiss army knife of Snakemaker, offering complete flexibility.  It is implemented by feeding the extension's state (settings, recorded history, opened screens, internal data structures) and README to the LLM, which is instructed to use a set of commands by outputting given URIs. These URIs are parsed by Snakemaker and the action performed. Actions allow to modify the extension settings, modify the recorded history, print new rules, modify the constructed DAG or the generated code during the Notebooks conversion.

\subsection{\bf LLM Interaction}

To ensure wide compatibility, interactions with LLMs are purely based on raw text. LLM prompts are dynamically generated from the current settings and data. Prompts are designed around the guidelines provided by the Whitepaper on Prompt Engineering \cite{G} and the widely known guidelines from OpenAI \cite{H}. Prompts are as compact and clear as possible, with a preference for positive instructions; context is provided in well-divided paragraphs in the upper part and the desired output format is described at the end. 

Prompts have then been fine-tuned iteratively, testing the pipeline, identifying frequent problems and preventing them. Some models, especially smaller ones, clearly suffer from limited knowledge of the Snakemake language and tend to repeatedly produce syntax errors such as using the dot notation instead of the pythonic dictionary access syntax. These common errors are largely prevented by using few-shot prompts or simply including precise instructions to avoid them. To keep prompts’ sizes limited, some frequent errors are fixed in a post-processing step using regex or small finite-state automata. The response format is, in most cases, JSON strings, allowing the extension to parse rich data structures from raw text responses. The stochastic nature of LLMs renders this process prone to parsing errors. Similarly to rule generation, common errors are identified and prevented with specific instructions, and the open-source tool Json-Repair \cite{jsonrepair} is employed as post-processing. In case of failure, an iterative approach is employed, similar to rule validation and correction.

\section{\bf Results}
\label{sec:RESULTS}

\begin{figure}[h]
\vspace{3mm}
 \begin{center}
 \includegraphics[width=0.8\textwidth]{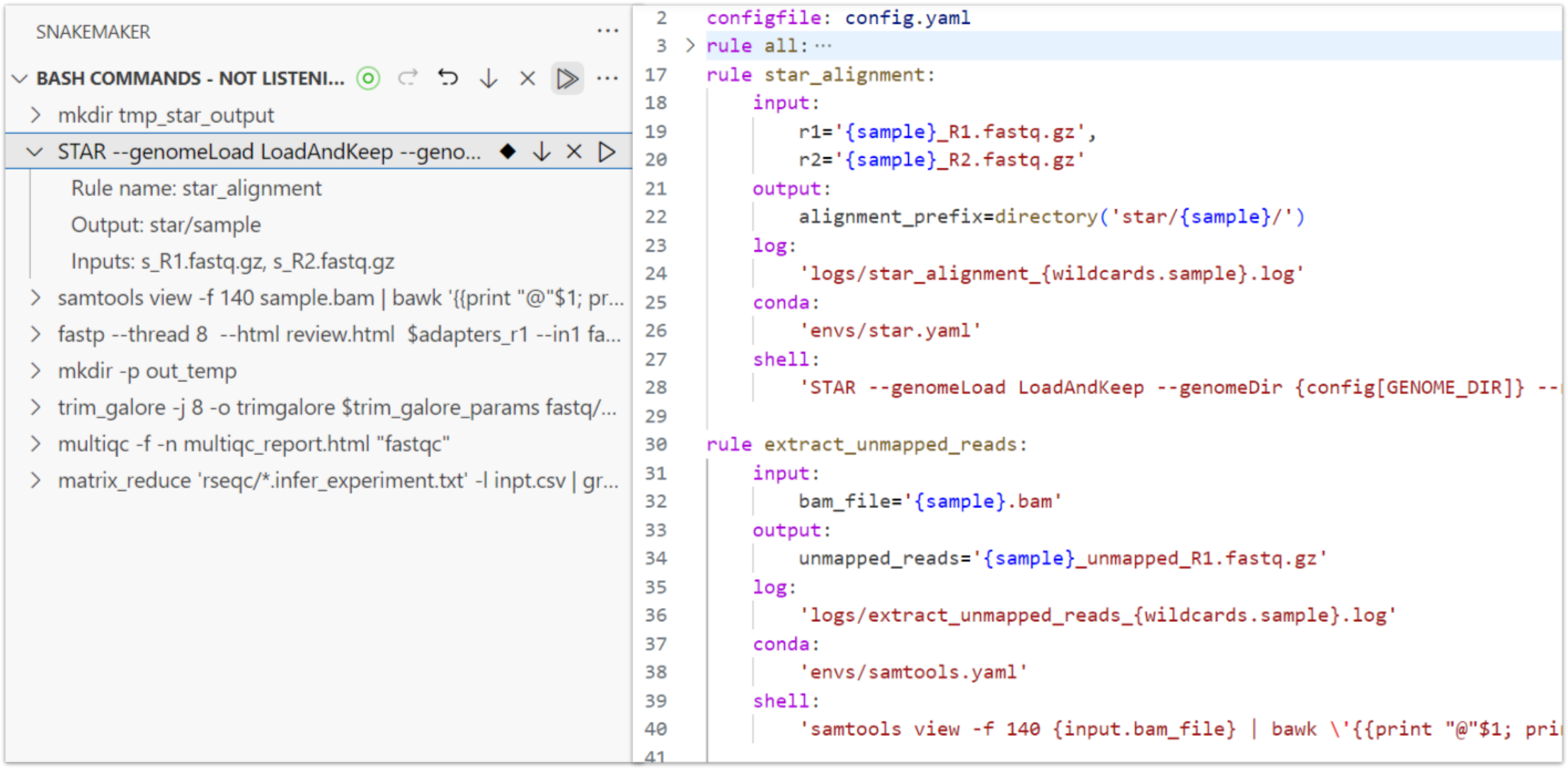}
\caption{\textbf{Screenshot from Snakemaker GUI}. Bash conversion feature.
\label{fig:screen}}
 \end{center}
\vspace{-8mm}
\end{figure}

Snakemaker allows to build Snakemake workflows from the recorded activity of the integrated bash terminal and from Ipython Notebooks. 

The bash feature runs non-invasively in the background, freeing researchers from manually tracking their work and environments. Commands are recorded, processed, and stored locally in a command history,  which is displayed in a graphical interface, where users can edit, delete, or group commands into composite entries. Tests with different LLMs show that they are reliable at distinguishing workflow-relevant commands from incidental ones (e.g. directory navigation, git interaction), with accuracy improving through user feedback. 

Users can trace back and convert the entire history or individual commands with a simple click. If a workflow already exists, the new rules are merged into it. The generated Snakefiles follow best practices, using wildcards, conda environments, config files for adaptability, per-rule logs and documentation. 

For users who are not working within the integrated terminal in VSCode, Snakemaker also supports manual input of previously executed commands. This can be done by copying the output of the shell’s built-in history command — a standard feature available in most Unix-like command-line interfaces. This command lists all previously executed commands in chronological order, providing a textual record of the user activity in the shell. While this approach allows the reconstruction of workflows outside the integrated environment, it lacks the richer metadata (like directory context, or environment information) that Snakemaker is able to capture automatically when run within the integrated terminal.

IPython notebooks, while great for prototyping, often become unwieldy. Snakemaker disassembles them into modular scripts with clear dependencies, connected by a Snakemake pipeline, resolving their global states into independent code chunks using regex and LLMs. 

The user has control over the process. In the first step a direct acyclic graph (DAG) describing the dependencies among notebook cells is computed and displayed graphically to the user, who can review it, spot errors, modify the \textit{read} and \textit{write} sets of each block, mark dependencies to be resolved with \textit{wildcards}, merge, split or delete cells. The user can also decide if a cell will be managed as a \textit{rule} or \textit{script} (see section \ref{sec:DATA-NOTEBOOK-EXPORT} for more details). In the following step, the user can manually adjust the generated code, and the changes are automatically propagated to downstream cells in the DAG. For example, modifying the format of an output file in a Snakemake rule results in the rewriting of the suffix code of the cell producing the output and the prefix blocks of all the cells that read it. 

Moreover, Snakemaker provides a chat assistant that can perform arbitrary actions requested by the user through natural language prompts. It allows for greater flexibility in the usage of the extension: in the bash feature, the user can request rules with custom prompts, covering user-specific needs, perform batch changes to the recorded history, explain or modify the current settings. The rules generated by the chat assistant can be directly appended to the workflow or fed to the pipeline defined in \ref{sec:DATA-SHELL-TRACKING} starting from step 3 for post-processing.  Since the notebook conversion feature requires more user input than the bash functionality, the flexibility provided by the chat assistant is particularly useful in this case. The assistant can review the DAG being built or the generated code, spot and fix errors and perform changes. The generated code is inevitably opinionated, as decisions must be made for the file formats and names, and the assistant allows to easily modify these decisions. 

Snakemaker currently requires LLMs of a certain size to perform effectively. Tests have shown optimal performance using state of the art models such as OpenAI 4o, suboptimal performance with open-source models under 70B parameters, and near-unusable results with 4B–8B parameter models, making local deployment challenging. However, its integration with GitHub Copilot lowers the barrier to accessing commercial, high-end models. The support for OpenAI-compatible APIs enables seamless connection to virtually any API provider. Settings of the extension allow users to disable individual features such as configuration generation, iterative validation, or workflow context inclusion, tailoring performance to different models strengths, speeds, and API costs. 

\section{\bf Conclusion}
\label{sec:CONCLUSIONS}

Snakemaker offers assistance in the development of sustainable data analysis pipelines while being non-intrusive into user workflows. It reduces the effort required to convert prototype code (level 0) into well-defined and organized pipelines (level 2). It does so by embracing the Snakemake language and its best practices, and employing LLMs as well as traditional programming algorithms to render the process semi-automatic. 

Being a VSCode extension, Snakemaker can integrate into the work environment of the users, and take advantage of the VSCode language API to connect to GitHub copilot, which is free for students and academics and a cost-effective option for most use cases, facilitating LLMs usage inside individuals workflows.

A potential area for future improvement is optimizing Snakemaker for the effective use of local LLM deployments. This could involve employing Retrieval-Augmented Generation (RAG) techniques with dedicated Snakemake language specifications, tutorials, and curated examples to enhance model performance even on smaller, locally hosted models.

Snakemaker lowers the barrier to structured workflows, encouraging a shift from ad-hoc analyses to sustainable, reproducible practices — advancing not just reproducibility, but a long-term culture of open, high-quality data analysis.

\section*{\bf Conflict of interests}
\label{sec:CONFLICT-OF-INTERESTS}
None declared. 

\section*{\bf Funding}
\label{sec:FUNDING}

This work has been supported by grant 2024-342822 (5022) GB-1609971 from the program Essential Open Source Software for Science Program of Chan Zuckerberg Initiative DAF, an advised fund of Silicon Valley Community Foundation. 

\section*{\bf Availability of data and software code}
\label{sec:AVAILABILITY}
Our software code is available at the following URL: https://github.com/molinerisLab/snkmaker 

\footnotesize
\bibliographystyle{unsrt}
\bibliography{bibliography_CIBB_file.bib} 
\normalsize

\end{document}